\documentclass[journal=jacsat,manuscript=article,layout=twocolumn]{achemso}

\usepackage{amssymb}
\usepackage{graphicx}
\usepackage{dcolumn}
\usepackage[version=3]{mhchem}
\usepackage{textcomp, gensymb}
\usepackage[fontsize=10pt]{fontsize}

\usepackage{xr}
\makeatletter
\newcommand*{\addFileDependency}[1]{
  \typeout{(#1)}
  \@addtofilelist{#1}
  \IfFileExists{#1}{}{\typeout{No file #1.}}
}
\makeatother

\newcommand*{\myexternaldocument}[1]{%
    \externaldocument{#1}%
    \addFileDependency{#1.tex}%
    \addFileDependency{#1.aux}%
}

\myexternaldocument{SIv3}

\author{Christian P. N. Tanner}
\affiliation{Department of Chemistry, University of California, Berkeley, CA 94720, USA}
\altaffiliation{Contributed equally to this work}

\author{Vivian R. K. Wall}
\affiliation{Department of Chemistry, University of California, Berkeley, CA 94720, USA}
\altaffiliation{Contributed equally to this work}

\author{Joshua Portner}
\affiliation{Department of Chemistry, James Franck Institute, and Pritzker School of Molecular Engineering, University of Chicago, Chicago, IL 60637, USA}

\author{Ahhyun Jeong}
\affiliation{Department of Chemistry, James Franck Institute, and Pritzker School of Molecular Engineering, University of Chicago, Chicago, IL 60637, USA}

\author{Avishek Das}
\affiliation{Department of Chemistry, University of California, Berkeley, CA 94720, USA}
\alsoaffiliation{Present Address: AMOLF, Science Park 104, 1098 XG, Amsterdam, The Netherlands}

\author{James K. Utterback}
\affiliation{Department of Chemistry, University of California, Berkeley, CA 94720, USA}
\alsoaffiliation{Present Address: Sorbonne Université, CNRS, Institut des NanoSciences de Paris, INSP, 75005 Paris, France}

\author{Leo M. Hamerlynck}
\affiliation{Department of Chemistry, University of California, Berkeley, CA 94720, USA}

\author{Jonathan G. Raybin}
\affiliation{Department of Chemistry, University of California, Berkeley, CA 94720, USA}

\author{Matthew J. Hurley}
\affiliation{Department of Physics, Arizona State University, Tempe, AZ 85287, USA}
\alsoaffiliation{Present Address: Department of Physics, Stanford University, Stanford, CA 94305, USA}

\author{Nicholas Leonard}
\affiliation{Department of Physics, Arizona State University, Tempe, AZ 85287, USA}

\author{Rebecca B. Wai}
\affiliation{Department of Chemistry, University of California, Berkeley, CA 94720, USA}

\author{Jenna A. Tan}
\affiliation{Department of Chemistry, University of California, Berkeley, CA 94720, USA}

\author{Mumtaz Gababa}
\affiliation{Department of Chemistry, University of California, Berkeley, CA 94720, USA}

\author{Chenhui Zhu}
\affiliation{Advanced Light Source, Lawrence Berkeley National Laboratory, Berkeley, CA 94720, USA}

\author{Eric Schaible}
\affiliation{Advanced Light Source, Lawrence Berkeley National Laboratory, Berkeley, CA 94720, USA}

\author{Christopher J. Tassone}
\affiliation{Stanford Synchrotron Radiation Lightsource, SLAC National Accelerator Laboratory, Menlo Park, CA 94025, USA}

\author{David T. Limmer}
\affiliation{Department of Chemistry, University of California, Berkeley, CA 94720, USA}
\alsoaffiliation{Chemical Sciences and Materials Sciences Divisions, Lawrence Berkeley National Laboratory, Berkeley, CA 94720, USA}
\alsoaffiliation{Kavli Energy NanoSciences Institute, University of California, Berkeley, CA 94720, USA}

\author{Samuel W. Teitelbaum}
\affiliation{Department of Physics, Arizona State University, Tempe, AZ 85287, USA}

\author{Dmitri V. Talapin}
\affiliation{Department of Chemistry, James Franck Institute, and Pritzker School of Molecular Engineering, University of Chicago, Chicago, IL 60637, USA}
\alsoaffiliation{Center for Nanoscale Materials, Argonne National Laboratory, Argonne, IL 60517, USA}

\author{Naomi S. Ginsberg}
\email{nsginsberg@berkeley.edu}
\affiliation{Department of Chemistry, University of California, Berkeley, CA 94720, USA}
\alsoaffiliation{Department of Physics, University of California, Berkeley, CA 94720, USA}
\alsoaffiliation{Molecular Biophysics and Integrated Bioimaging Division, Lawrence Berkeley National Laboratory, Berkeley, CA 94720, USA}
\alsoaffiliation{Materials Sciences and Chemical Sciences Division, Lawrence Berkeley National Laboratory, Berkeley, CA 94720, USA}
\alsoaffiliation{Kavli Energy NanoSciences Institute, University of California, Berkeley, CA 94720, USA}
\alsoaffiliation{STROBE, NSF Science \& Technology Center, Berkeley, CA 94720, USA}

\title{Enhancing nanoscale charged colloid crystallization near a metastable liquid binodal}

\begin{document}

\begin{abstract}

Achieving predictive control over crystallization using non-classical nucleation while avoiding kinetic traps is a roadblock toward designing materials with new functionalities. We address these challenges by inducing bottom-up assembly of nanocrystals (NCs) into ordered arrays, or superlattices (SLs). Using electrostatics, rather than density, to tune the interactions between particles, we watch self-assembly proceeding through a metastable liquid phase. We systematically investigate the phase behavior as a function of quench conditions \textit{in situ} and in real time using small angle X-ray scattering (SAXS). Through quantitative fitting to colloid, liquid, and SL models, we extract the time evolution of each phase and the system phase diagram, which we find to be consistent with  short-range attractive interactions. Using the phase diagram's predictive power, we establish control of the self-assembly rate over three orders of magnitude, and identify
one- and two-step self-assembly regimes, with only the latter implicating the metastable liquid as an intermediate. Importantly, the presence of the metastable liquid  increases SL formation rates relative to the equivalent one-step pathway, and SL order counterintuitively increases with the rate, revealing a highly desirable and generalizable kinetic strategy to promote and enhance ordered assembly.

\end{abstract}

\subsubsection{Keywords:}
nanocrystals, 
self-assembly,
crystallization, 
X-ray scattering, 
\textit{in situ} measurement,
non-classical nucleation
\subsubsection{}
\maketitle

\begin{figure*}
\includegraphics[width=16cm]{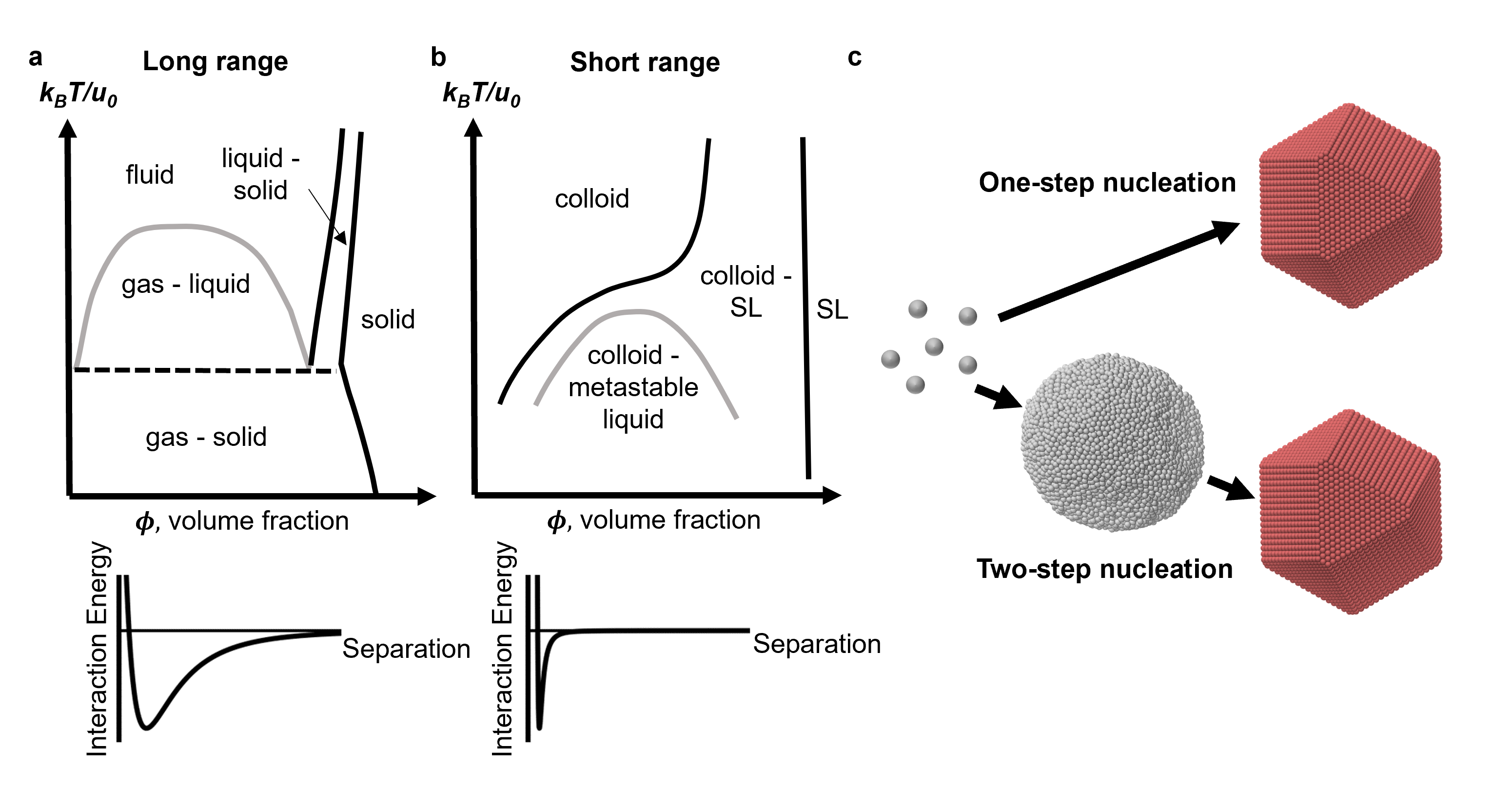}
\caption{Impact of interparticle potentials on thermodynamics and kinetics of self-assembly. a. Effective temperature ($k_\mathrm{B}T/u_0$) vs. volume fraction ($\phi$) phase diagram (top) and sketch of interaction potential vs. interparticle separation (bottom) for particles interacting via long-range interparticle potentials. b. Sketch of effective temperature vs. volume fraction phase diagram (top) and sketch of interaction potential vs. interparticle separation (bottom) for particles interacting via short-range interparticle potentials. In both a and b, $u_0$ is the magnitude of the minimum value of the interaction potential. c. Schematic of one-step (top) and two-step (bottom) SL formation pathways.}
\label{Fintro}
\end{figure*}

The range of interparticle interactions relative to the size of the interacting particles determines the qualitative features of a system’s equilibrium phase diagram\cite{ten_wolde_enhancement_1997,noro_extended_2000,haxton_crystallization_2015}. Atomic systems have a long range of the interactions relative to atomic scales, yielding temperature-density phase diagrams like in \textbf{Figure \ref{Fintro}a}\cite{barroso_solidfluid_2002,kofke_direct_1993}. As this range is decreased, liquid phase stability diminishes and the gas-liquid binodal or, for a colloidal system, the colloid-liquid binodal (grey) shifts to lower temperature. The colloid-liquid binodal is a fluid-fluid binodal in which the colloidal phase is the low-density fluid and the liquid phase is the high-density fluid composed of densely packed colloidal particles. At a range of less than $\sim$20\% of the particle diameter\cite{noro_extended_2000}, 
the colloid-liquid binodal (\textbf{Figure \ref{Fintro}b}, grey curve) sits entirely below the colloid-solid binodal (\textbf{Figure \ref{Fintro}b}, black curve), rendering the liquid phase metastable \cite{ten_wolde_enhancement_1997,noro_extended_2000,hagen_determination_1994}. Consequently, as an alternative to typical one-step crystallization directly from the colloidal phase (\textbf{Figure \ref{Fintro}c} top), systems like proteins with short range attractive interactions may crystallize via a metastable liquid intermediate (\textbf{Figure \ref{Fintro}c} bottom)\cite{ten_wolde_enhancement_1997,wedekind_optimization_2015,haxton_crystallization_2015,savage_experimental_2009,zhang_charge-controlled_2012,lee_entropic_2019,du_non-classical_2024}. 
For proteins and in simulations\cite{ten_wolde_enhancement_1997,wedekind_optimization_2015,haxton_crystallization_2015,haas_interface_2000,zhang_charge-controlled_2012,galkin_control_2000,du_non-classical_2024}, the two-step crystallization pathway has been shown to speed up crystallization over the equivalent one-step rate without increasing the crystallization driving force since the high density of particles in the liquid phase facilitates the order fluctuations necessary to form the crystal. Many questions, however, remain about the generality and underlying mechanisms of two-step crystallization. In particular, as seen in colloid-polymer mixtures and nanocrystal (NC) systems with similar interaction ranges to many protein systems, gelation often occurs before a metastable liquid can form. 
How subtle differences in the interparticle interactions alter crystallization outcomes is therefore essential to elucidate so that design principles can leverage the metastable liquid available via short-range interactions\cite{soga_metastable_1999,noro_role_1999,kpoon_gelation_1995,du_non-classical_2024}. Recently, colloid-metastable liquid phase separation as a precursor to crystallization has been seen in electrostatically stabilized PbS semiconductor NCs dispersed in electrolyte solutions, \cite{coropceanu_self-assembly_2022} and the fluidity of the metastable liquid state in the same system has been confirmed via X-ray photon correlation spectroscopy.\cite{tanner2024origins} To fully uncover the crystallization mechanism and the role of underlying interparticle interactions requires systematic characterization of self-assembly trajectories in many different regimes.

Measuring the phase evolution of nanoscale systems can be challenging due to the fast time- and small length-scales intrinsic to them. Typically, the equilibrium phase behavior of colloids is measured using optical microscopy and scattering methods\cite{pusey_phase_1986,gasser_real-space_2001,dinsmore_phase_1995,verhaegh_fluid-fluid_1996}, yet optical methods insufficiently capture sub-diffraction nanoscale evolution. Recent small angle X-ray scattering (SAXS) has  measured the evolving phase coexistence, \textit{in situ} and in real-time, as colloidal NCs convert to superlattice (SL),\cite{tanner_situ_2024,weidman_kinetics_2016,korgel_small-angle_1999,lu_resolving_2012,lokteva_real-time_2021,wu_high-temperature_2017,abecassis_gold_2008,marino_temperature-controlled_2023,grote_x-ray_2021,josten_superlattice_2017,geuchies_situ_2016,qiao_situ_2023} an ordered solid of NCs\cite{boles_self-assembly_2016,murray_self-organization_1995,shevchenko_structural_2006,smith_self-assembled_2009,bian_shape-anisotropy_2011}. Compared to visible light, X-rays provide more detailed, nanoscale information that is especially useful to distinguish different phases that do not separate on a macroscopic length scale on self-assembly time scales. Distinguishing coexisting colloidal, liquid, and solid phases remains challenging, however, due to their overlapping features  in reciprocal space, limiting qualitative deductions from scattering patterns and common quantitative Bragg peak analyses\cite{warren_x-ray_1990}. Special care is therefore needed to accurately extract the phase behavior of NC systems as well as the kinetics of their phase transformations from SAXS data. 

Despite the fast time- and short length-scales of NC phase behavior, as well as the challenge of distinguishing phases in SAXS data, we systematically characterize the self-assembly of SLs from electrostatically stabilized PbS NCs using \textit{in situ} SAXS, capturing multiple kinetic pathways, including the evolution of the liquid-state intermediate. To do so, we leverage the gas-tight reactor that we recently developed to study Au NC SL self-assembly\cite{tanner_situ_2024}. We developed a model to fit SAXS patterns that quantitatively distinguishes between colloidal, liquid, and SL phases. From these fits we obtain the evolution of each phase as a function of time and quench depth ($k_\mathrm{B}T/u_0$, where $u_0$ is the depth of the interparticle potential) to deduce self-assembly mechanisms and kinetic rates that are not retrievable \textit{ex situ}. By electrostatically tuning interparticle interactions, we reliably control self-assembly rates over three orders of magnitude as well as the mechanism: crystallization of SLs occurs either directly from the colloidal phase or in a two-step process with a metastable liquid intermediate. Our results suggest that the two-step pathway increases crystallization rates over the equivalent one-step pathway \textit{while simultaneously increasing SL order}, a highly attractive kinetic strategy. Through comparison of the  PbS NC system to its Au NC counterpart,\cite{tanner_situ_2024} we clarify how specific chemical and physical properties impact interparticle interactions and self-assembly, identifying the balance of interactions between the PbS NCs that avails the liquid phase and control of crystallization. In addition to establishing these design principles, our experimental and analysis methods should reveal the kinetic and thermodynamic complexity of a wide range of other nanosystems at the forefront of functional materials design and provide a strategy to achieve similar order in place of gelation with larger scale colloids.\cite{boles_self-assembly_2016,murray_self-organization_1995,shevchenko_structural_2006,smith_self-assembled_2009,bian_shape-anisotropy_2011,haxton_crystallization_2015,campbell_dynamical_2005,tsurusawa_hierarchical_2023}.


\textbf{\textit{Phases and phase evolution of self-assembling short-range-interacting nanocrystals.}} To determine the effect of a metastable liquid phase on the self-assembly of NC SLs, we study 5.8$\pm$0.3 nm diameter electrostatically stabilized PbS NCs (\textbf{Figure S\ref{SI-char}a-c}) dispersed in a polar solvent (N-methylformamide (NMF) mixed with N,N-dimethylformamide (DMF)), with multivalent, inorganic thiostannate ($\mathrm{Sn}_2\mathrm{S}_6^{4-}$) ligands on their surface\cite{coropceanu_self-assembly_2022}.
We quench to a condensed phase (\textbf{Figure S\ref{SI-char}d}) by injecting a solution of additional multivalent salt ($\mathrm{K}_3\mathrm{AsS}_4$)  into NC solutions of varying volume fraction. The salt  screens the initial electrostatic repulsion generated by the charged thiostannate ligands to achieve attractive interactions between NCs. Over tens of single-shot experiments we systematically tune the solution ionic strength (see \textbf{Methods}) from 0.9 to 1.5 M to vary the quench depth, i.e., driving force for self-assembly, and we tune the volume fraction of NCs from $\phi_0\sim 3\times10^{-5}$ to $8 \times 10^{-3}$. 
We thus traverse a two-dimensional  (2D) parameter space like those in the phase diagrams in \textbf{Figures \ref{Fintro}a,b}.

To follow the entire self-assembly process \textit{in situ}, we employ a gas-tight reactor previously developed for SAXS\cite{tanner_situ_2024}. A quartz cuvette with thin, X-ray-transparent windows initially contains NCs  colloidally dispersed in an NMF/DMF mixture (\textbf{Figure S\ref{SI-aparatus}}). The cuvette is connected via tubing to a syringe on a syringe pump that injects a controlled amount of $\mathrm{K}_3\mathrm{AsS}_4$ in NMF, inducing self-assembly. A magnetic stir bar in the cuvette ensures the X-rays  sample the solution  representatively. We collect 2D SAXS detector images before, during, and after salt injection with down-to-millisecond time resolution and for up to two hours post-injection. By azimuthally averaging detector images, we obtain one-dimensional SAXS patterns, $I(q)$, the scattered intensity as a function of  momentum transfer, $q$. The time-evolution of $I(q)$ reveals the detailed evolution of the system's coexisting phases, including the liquid phase density and SL crystallinity.

\begin{figure*}
\includegraphics[width=12cm]{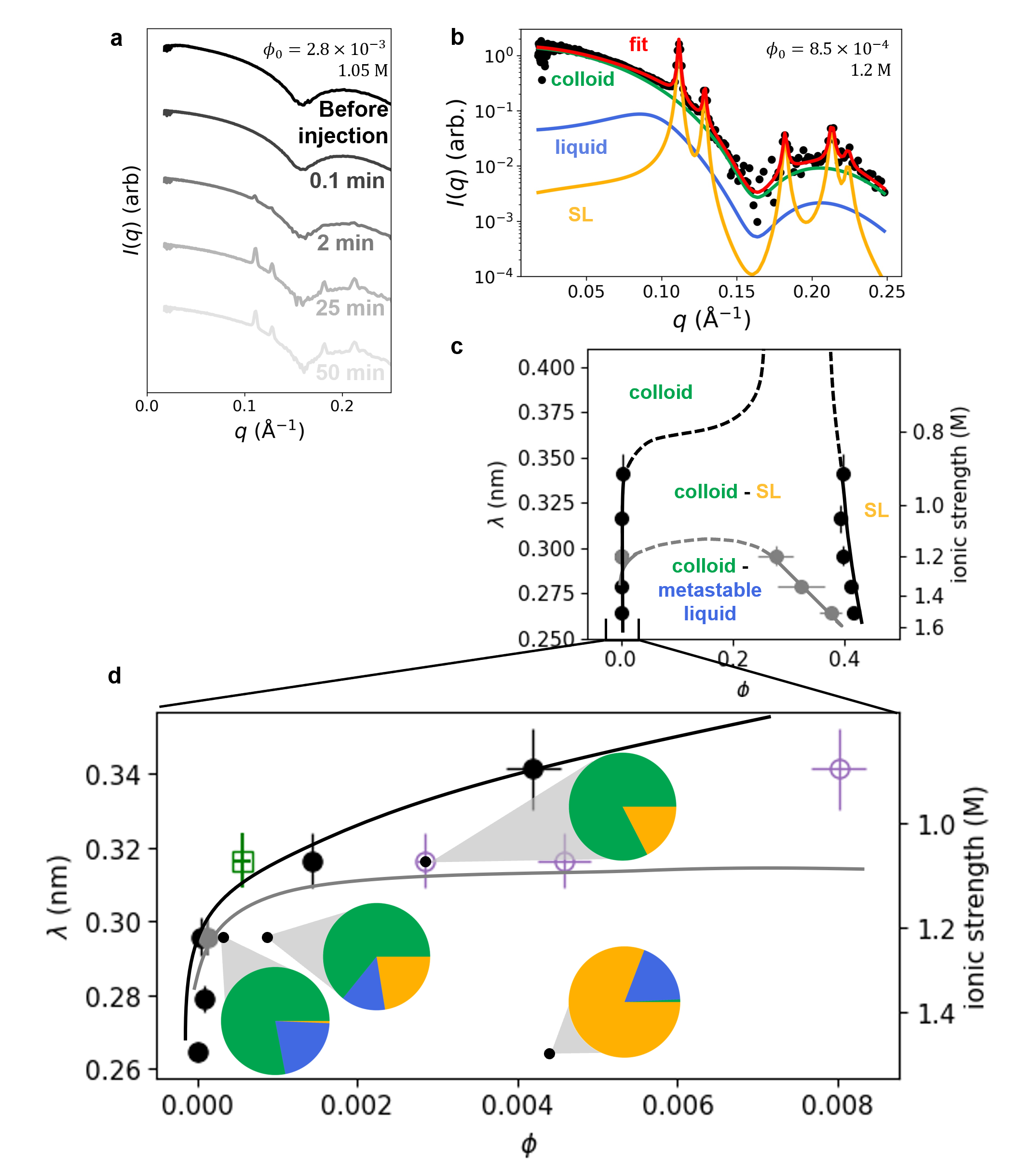}
\caption{Quantitative analysis of quench- and volume fraction-dependent SAXS patterns. a. One-dimensional SAXS patterns as a function of time after injection of excess salt for an example experiment using NC volume fraction $\phi_0 = 2.8 \times 10^{-3}$ and a solution ionic strength (quench depth) of 1.05 M. Additional examples at other volume fractions and quench depths can be found in \textbf{Figure S\ref{SI-waterfall}}. b. Quantitative fit (red) of SAXS pattern obtained using colloidal NC (green), liquid (blue), and SL (yellow) phases at the same NC volume fraction and quench depth as in a after equilibrating $>$ 1 hour post quench for an experiment at $\phi_0 = 8.5 \times 10^{-4}$ and a solution ionic strength  of 1.2 M. Additional examples of fits corresponding to the four cases in \textbf{Figure S\ref{SI-waterfall}} can be found in \textbf{Figure S\ref{SI-fits}}.  c. Quantitative phase diagram for electrostatically stabilized PbS NCs obtained from experimental observations. Black  points fall on the colloid-SL binodal and grey  points fall on the colloid-liquid binodal. We sketched the curves between  $\phi\sim$ 0.01 and $\phi\sim$ 0.3 with dashing as an interpolation between the solid curves dictated by experimental data; the dashed curves are not meant to be quantitative. Vertical error bars indicate the standard deviations of the Debye lengths, $\lambda$, of the solutions based on the uncertainty of the volume and concentration of the injected salt solution. Horizontal error bars indicate the standard deviations of the colloidal, liquid, and SL volume fractions due to the same volume uncertainty and uncertainty from SAXS fitting. Black (colloid-SL) and grey (colloid-liquid) phase boundary curves are sketched as a visual guide. d. Close-up of low-density region of the phase diagram with additional points included. The open green squares indicate ($\phi,\lambda$) phase diagram locations in which the system is found only in the colloidal state. The open purple circles indicate ($\phi,\lambda$) locations in which no liquid component was required in the model to fit the SAXS patterns. The relative fractions of the different phases observed are represented for four distinct regimes whose fits are shown in \textbf{Figure S\ref{SI-fits}} as pie charts color coded to match the fits in a (colloidal NC: green, liquid: blue, SL: yellow).}
\label{FSAXS}
\end{figure*}

\textbf{Figure \ref{FSAXS}a} shows an example of a  time lapse with key time points over the course of an hour.  Before salt injection the SAXS patterns show the decaying oscillatory scattering characteristic of 5.8 nm colloidal NCs (top curve). After the salt injection, the first five Bragg peaks of an \textit{fcc} SL appear, e.g., (111) at $q\sim$0.11 $\text{\r{A}}^{-1}$, (200) at 0.12 $\text{\r{A}}^{-1}$, growing progressively  until the system approaches equilibrium. Additional examples with distinct combinations of NC volume fractions and quench depths are found in \textbf{Figure S\ref{SI-waterfall}} and \textbf{Figure S\ref{SI-noSL}}. Globally, Bragg peaks emerge more quickly at deeper quenches and at higher NC volume fractions, as expected. The Bragg peaks are broader at deeper quenches, indicating that at deeper quenches the SL domains are smaller and more disordered.\cite{warren_x-ray_1990}. While the colloidal form factor and SL Bragg peaks are discernible by eye, noting the absence or presence of a liquid phase  is not. 

To adequately describe the SAXS patterns over a range of NC concentrations and ionic strengths, we expanded a model that we previously developed\cite{tanner_situ_2024} to fit SAXS patterns of systems containing colloidal and SL phases in coexistence by including an additional term for the scattering from a NC liquid phase. We model the background-subtracted scattered intensity as $I(q) = I_{\mathrm{colloid}}(q) + I_{\mathrm{liquid}}(q) + I_{\mathrm{SL}}(q) + I_{\mathrm{LQ}}(q)$ (see \textbf{Methods, Figures S\ref{SI-background}, S\ref{SI-sims}a-c} and associated text). Here, $I_{\mathrm{colloid}}(q)$ is the scattered intensity from dilute, polydisperse hard spheres, $I_{\mathrm{liquid}}(q)$ is the scattered intensity from a liquid of hard spheres at a volume fraction $\phi_\mathrm{liquid}$, $I_{\mathrm{SL}}(q)$ is the scattered intensity from finite-sized \textit{fcc} SLs, and $I_{\mathrm{LQ}}(q)$ is the scattered intensity at low $q$ due to the finite size of high-density structures formed.  The model assumes that the different components are statistically uncorrelated, which we validated on simulated systems (see \textbf{SI}, \textbf{Figure S\ref{SI-sims}}). We find that the model fits the experimental data well (see \textbf{Figures S\ref{SI-resid}, S\ref{SI-covar}, S\ref{SI-NMF}}) at all time points, NC volume fractions, and quench depths and also provides detailed, instantaneous information about each phase, such as the SL crystallinity and liquid phase density. Examples are shown near equilibrium in \textbf{Figure \ref{FSAXS}b} and \textbf{Figure S\ref{SI-fits}}. 

Depending on the NC volume fraction and quench depth, we find cases in which the system contains strictly colloid and SL in coexistence (\textbf{Figure S\ref{SI-fits}a}) or cases for which the colloid coexists with the SL and liquid phases (\textbf{Figures \ref{FSAXS}b, S\ref{SI-fits}b-d}). To consolidate all of the long-time equilibrium behavior of the system as a function of the NC volume fraction and quench depth parameters explored, we indicate experimentally-determined points on the binodal curves that delineate the phase boundaries of this system's phase diagram in \textbf{Figure \ref{FSAXS}c} (see \textbf{Methods}). These points specify the volume fraction, or density, of NCs in the colloidal or SL phases as a function of quench depth within the binodal curves. We cannot explicitly know a potential well depth for the yet-unspecified short-range potential of the quenched NCs. Thus, even though the NC solvation has characteristics that go beyond a simple double layer,\cite{silvera_batista_nonadditivity_2015,zhang_stable_2017,kamysbayev_nanocrystals_2019} we use the Debye length, $\lambda$, of the solution as a figure of merit for the quench depth on the vertical axis. The Debye length combines information about the varying ionic strengths of the solution and the (fixed) dielectric constant of the solvent. Although this choice does not incorporate the impact of the steric and van der Waals contributions from the NCs into the quench depth parameter, these forces should not vary from quench to quench since the same NC stock solution was used for all measurements. For clarity, the ionic strength is also indicated on the right-hand side of the plot. 

To determine the points at each quench depth on the low-density side of the phase diagram in \textbf{Figure \ref{FSAXS}c}, i.e., the volume fraction of the colloidal phase, $\phi_\mathrm{colloid}$, we multiply the fraction of NCs remaining in the colloidal phase post-quench by the total volume fraction of NCs in the system. We calculate the high-density colloid-SL binodal points' locations from the corresponding densities of the SL phase, $\phi_\mathrm{SL}$, at each quench depth based on the  \textit{fcc} (111) Bragg peak positions. The volume fraction of the SL phase is lower than expected for typical \textit{fcc} crystals due to the presence of ligand and electrolyte molecules in the interstices of the lattice, neither of which is included in the computed NC volume. Here, we include an additional colloid-metastable liquid binodal (grey data points and curve in \textbf{Figure \ref{FSAXS}c}) that bounds the colloid-metastable liquid coexistence. The high-density side of this binodal is defined by the liquid volume fraction, $\phi_\mathrm{liquid}$, extracted from the $I_\mathrm{liquid}(q)$ component of the fits. The low-density side is obtained from the density of the colloidal phase in coexistence with the metastable liquid, which is  extracted from cases for which we observe a separation of time scales between liquid and SL formation (grey circle in \textbf{Figure \ref{FSAXS}d}). Limitations on our estimates of these binodals are discussion in the SI. In order to resolve the binodals between $\phi$ $\sim$ 0.01 and 0.3, higher total NC volume fractions 
that are difficult to stabilize would be required. We did not do so in this work due to the greater difficulty stabilizing the colloidal phase at high NC volume fractions. Nevertheless, experiments for which we injected insufficient salt to condense the colloid into either an ordered or disordered dense phase allow us to bound the colloid-SL binodal location at low volume fraction (open green squares in \textbf{Figure \ref{FSAXS}d}). Although the critical density is experimentally inaccessible, we constrain the location and height of the low-density colloid-metastable liquid binodal based on phase diagram locations that resulted in one-step SL formation pathways (open purple circles in \textbf{Figure \ref{FSAXS}d}). Another phase diagram obtained in the same way for a different stock solution of PbS NCs is shown in \textbf{Figure S\ref{SI-pd}}.

\begin{figure*}
\includegraphics[width=18cm]{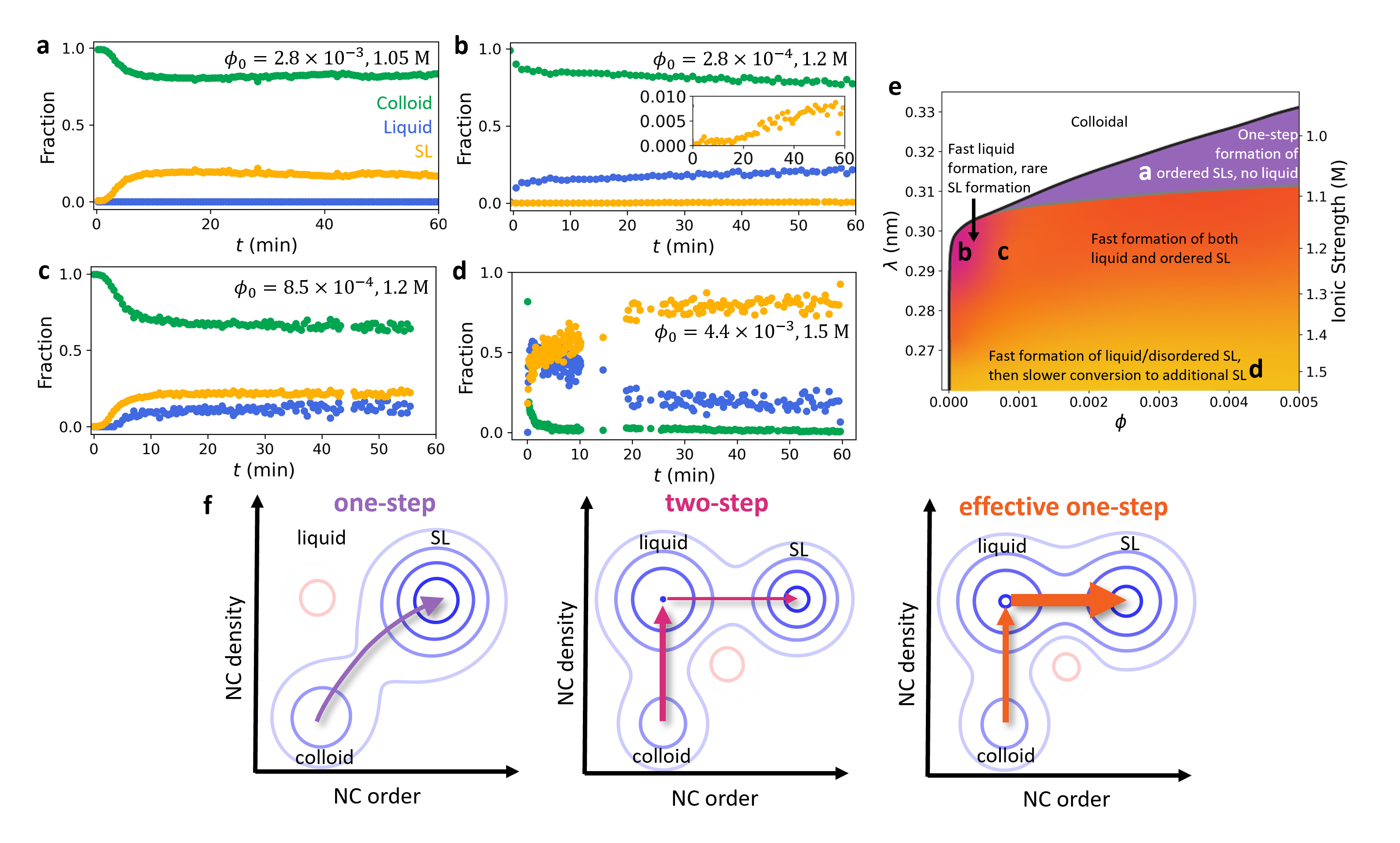}
\caption{Kinetic analysis of quench- and volume fraction-dependent SAXS. a-d. Fraction of NCs in the colloidal (green), liquid (blue), and SL (yellow) phases as a function of time for NC volume fractions and quench depth cases covered in \textbf{Figure S\ref{SI-waterfall}a-d} and \textbf{Figure S\ref{SI-fits}a-d} (see SI for further details). (The absence of data between $\sim$10 and $\sim$19 min in panel d is due to a pause incurred when changing detector sampling frequencies but does not compromise the interpretation of the kinetics for the quench conditions probed.) e. Map of experimentally explored parameter space, indicating approximate progression of regions exhibiting the four kinetic behaviors shown in a-d. Black and grey curves sketch the approximate colloid-SL and colloid-metastable liquid binodals, respectively, as in \textbf{Figure \ref{FSAXS}c}. f. Distinct crystallization pathways illustrated on cartoon free energy surfaces as a function of NC order and density. Arrow widths indicate relative rates within each pathway. Blue shades indicate negative relative free energies; red shades indicate positive relative free energies.} 
\label{Fkinetics}
\end{figure*}

In order to study the kinetics of self-assembly, we calculate and plot in \textbf{Figure \ref{Fkinetics}a-d} the time evolution of the fraction of NCs in each phase for four distinct classes of kinetic behaviors observed as a function of quench depth and NC volume fraction. The SAXS patterns and fits for these four experiments are shown in \textbf{Figure S\ref{SI-waterfall}a-d} and \textbf{Figure S\ref{SI-fits}a-d}. 
In \textbf{Figure \ref{Fkinetics}a} the SL phase arises over minutes and plateaus in the absence of a liquid phase; in \textbf{Figure \ref{Fkinetics}b} the liquid phase arises and plateaus within seconds, but the SL phase does not appear until $\sim$30 min; in \textbf{Figure \ref{Fkinetics}c} the liquid and SL phases arise over several minutes and plateau concurrently; and in \textbf{Figure \ref{Fkinetics}d} the liquid and SL phases arise concurrently within 0.1 min, but the SL phase subsequently grows at the expense of the liquid phase in the following several minutes.

\textbf{\textit{Deducing features of electrostatically tunable nanoscale interparticle potentials.}} Having described our SAXS data and analysis, we now discuss the implications of our findings in terms of inferred interaction potentials, advantages of SL formation pathways, and emerging design principles. Previously, the phase behavior of these NCs was determined qualitatively via observing macroscopic phase separation in cuvettes at long times post-quench\cite{coropceanu_self-assembly_2022}. Here, we go beyond this initial work by directly and quantitatively measuring the phase diagram of these NCs over a large range of volume fractions and quench depths
(\textbf{Figure \ref{FSAXS}c,d}). Understanding the different contributions to the associated interparticle potentials and their dependence on external control parameters enables the fine tuning of interactions necessary to optimize ordered SL self-assembly for short-range interacting nanoscale particles. Both the equilibrium phase coexistence and kinetic mechanisms of SL self-assembly from electrostatically stabilized NCs are ultimately determined by the underlying depth and range of interactions between NCs. The range of interactions can be inferred from the extracted phase diagram, which is consistent with the phase diagrams of other particles interacting with short-range interparticle potentials. Based on the Noro-Frenkel law of corresponding states\cite{noro_extended_2000} and the effective size of the NCs ($\sim$6.4-7.8 nm, see \textbf{Methods}), we infer the surface-to-surface range of the interparticle potential to be 1.3-1.6 nm in order for the liquid phase to be metastable. Simulation studies bound
interaction well depths of short-range interacting systems to $2k_\mathrm{B}T<u_0<6k_\mathrm{B}T$ range\cite{haxton_crystallization_2015}. 

We further specify the interaction ranges and well depths by comparing the self-assembly of PbS NCs in this work with that of previously studied electrostatically stabilized Au NCs 
coated with the same thiostannate surface ligands. When Au NCs are quenched in the same way as in this work, they form SLs much more rapidly (within seconds), the SLs are more disordered, and  there is no evidence from \textit{in situ} measurements of a metastable liquid phase\cite{tanner_situ_2024,hurley2024situ}. The faster kinetics imply that Au NC interparticle potentials have larger values of $u_0$ than PbS NC interparticle potentials for similar quench conditions despite also being short-range. To explain the different kinetics using findings in Ref.\cite{haxton_crystallization_2015}, we propose that the PbS NC interparticle potentials likely have $u_0$$\sim$2-3 $k_\mathrm{B}T$ in order to generate highly ordered SLs, while the Au NC interparticle potentials likely have $u_0$$\sim$3-6 $k_\mathrm{B}T$. Such a difference would imply that, for the same change in external control parameter, such as solution ionic strength, the change in Au NC $u_0$  is larger than the change in PbS NC  $u_0$. This difference in sensitivity is likely caused by PbS NCs having weaker vdW attraction than Au NCs as well as the fact that solvents with a more strongly screening, higher dielectric constant were used for the PbS NC system compared to the Au NC system.\cite{tanner_situ_2024} The presence of a metastable liquid phase in the PbS NC SL self-assembly and the absence of  observed liquid in the Au NC case also suggests that, in combination with qualitative predictions from Derjaguin-Landau-Verwey-Overbeek (DLVO) theory\cite{israelachvili_intermolecular_2011}, Au NC interactions are even shorter in range with respect to the effective size of an Au NC than PbS NC interactions are with respect to their effective size (\textbf{Figure S\ref{SI-Au}}). This possibility could be due to Au NCs having larger effective sizes relative to the bare NC core than PbS NCs have due to their greater surface ligand coverage, which is expected due to their higher  NC-to-solvent dielectric constant ratio.\cite{guerrero_garcia_polarization_2014} In summary, while both systems have accessible short-range interaction regimes whose attractive wells are not so deep nor so short\cite{noro_role_1999} as to reach commonly-observed gelation,\cite{campbell_dynamical_2005,ruiz-franco_role_2021,tsurusawa_hierarchical_2023} we infer that PbS NC interactions are softer and longer in range than Au NC interactions when employing the same electrostatically stabilizing ligands and salts. The sensitivity of the PbS interparticle potentials to external control parameters ultimately enables the fine control necessary to tune the thermodynamic state, self-assembly pathway, and  kinetics in a nanoscale system.

\begin{figure*}
\includegraphics[width=9cm]{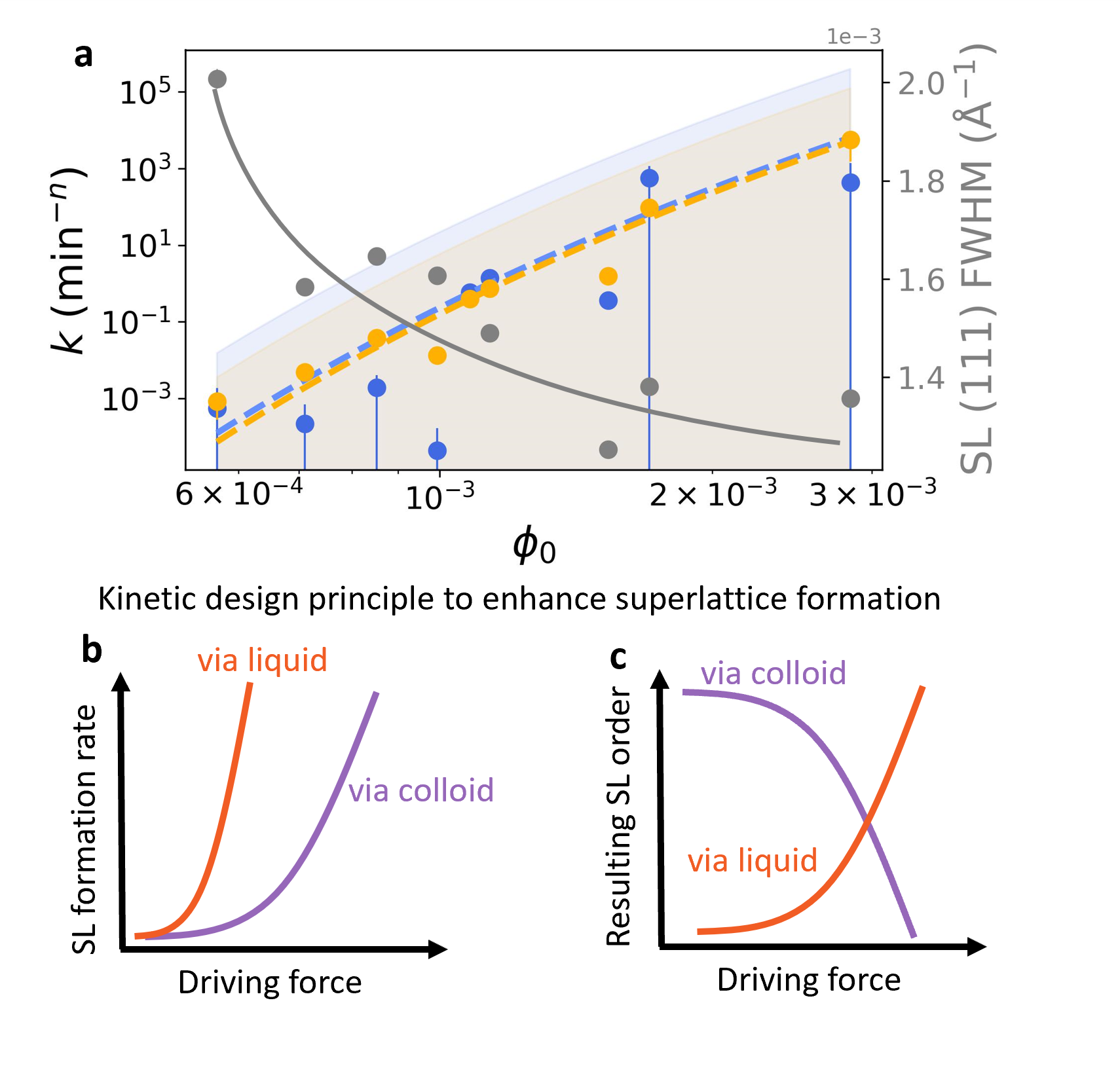}
\caption{Deduced design principles for optimal SL self-assembly. a. Avrami rate constants, $k$, for the liquid (blue) and SL (yellow) phases, and \textit{fcc} (111) Bragg peak full width at half maximum (FWHM) (grey) as a function of volume fraction, $\phi_0$, for a set of experiments within the orange region in \textbf{Figure \ref{Fkinetics}e}. The rate constants are fit to the functional dependence of $k$ on $\phi_0$ as predicted by classical nucleation theory (dashed curves) and the grey curve is a visual guide to the trend. 95$\%$ confidence intervals are shown for the rate constant model for each of the liquid and SL phases. Sketches of anticipated design principles to optimize SL formation rate (b) and resulting SL crystallinity (c) are shown as a function of driving force (quench depth). These design principles are anticipated to hold specifically for quenches to phase diagram locations below the colloid-metastable liquid binodal.}
\label{Fdesign}
\end{figure*}

\textbf{\textit{Liquid-containing kinetic pathways for control of nanoscale self-assembly.}} To further examine the different kinetic pathways available to achieve this fine control over self-assembly, \textbf{Figure \ref{Fkinetics}e} shows the region of the phase diagram in \textbf{Figure \ref{FSAXS}c,d} that we experimentally explored, with the approximate regions that exhibit the four distinct kinetic behaviors shown in \textbf{Figure \ref{Fkinetics}a-d} indicated. \textbf{Figure 3f} shows cartoon free energy surfaces illustrating the different crystallization pathways, where the pathways reflect the locations and relative scales of any free energy barriers between basins that are consistent with the corresponding thermodynamic states.
The SL growth in the absence of liquid observed in \textbf{Figure \ref{Fkinetics}a} suggests that, at a higher volume fraction and with a shallow quench, the SL forms directly from the colloidal phase (\textbf{Figure \ref{Fkinetics}e} purple, \textbf{Figure \ref{Fkinetics}f left}). The separation of time scales between the growth of the liquid and SL phases at lower volume fraction/moderate quench depth (\textbf{Figure \ref{Fkinetics}e} magenta and \textbf{Figure \ref{Fkinetics}b}) and the SL growth at the expense of the liquid at higher volume fraction/deep quench depth (\textbf{Figure \ref{Fkinetics}e} yellow and \textbf{Figure \ref{Fkinetics}d}) are consistent with a SL formation mechanism that includes a liquid intermediate (\textbf{Figure \ref{Fkinetics}f center}). Thus, our data suggest that electrostatically stabilized PbS NC SLs can form through both mechanisms in \textbf{Figure \ref{Fintro}c}. 

In \textbf{Figure \ref{Fkinetics}c}, the liquid and SL appear concurrently within experimental error, which is not formally expected from the one-step or the two-step pathways. 
This observation motivates an Avrami analysis on a series of $\sim$10 experiments at a constant quench depth of $\lambda$ = 0.296 nm (1.2 M) and varied volume fraction (\textbf{Figure \ref{Fkinetics}e} orange, \textbf{Figure S\ref{SI-Avrami}}) 
to describe SL nucleation and growth rates as a function of $\phi_0$ in this region of the parameter space (yellow in \textbf{Figure \ref{Fdesign}a}). 
 Interestingly, repeating this same procedure for the rate of emergence of the liquid phase vs $\phi_0$ reproduces an essentially similar trend (blue). This similarity is observed despite the more significant challenge of obtaining the rate constants for the more smoothly varying structure factor of the liquid, especially at low $\phi_0$ where the amplitude of this contribution to $I(q)$ is quite small. To combine the data from different volume fractions to obtain a much more robust assessment of the growth rates than can be obtained from a single Avrami fit on its own, we fit the Avrami rate constant, $k$ vs $\phi_0$ for the liquid and SL phases to a function arising from classical nucleation theory (dashed curves, see \textbf{Methods}).
Given these trends of Avrami rate constant $k$ vs. 
$\phi_0$ for the liquid and SL phases match reasonably well over several orders of magnitude (\textbf{Figure \ref{Fdesign}a}), we propose that kinetics in this region are one-step like in that they are limited by a transition from a colloidal to a dense phase, irrespective of its order. The colloid-to-liquid transition is rate-limiting, and the liquid to SL transition is fast, suggesting a free energy landscape such as that sketched at right in \textbf{Figure \ref{Fkinetics}f}.\cite{ten_wolde_enhancement_1997} We corroborated this finding with \textit{in situ} dynamic light scattering (DLS) measurements during self-assembly, which show that the average size of individual dense clusters grows as $\sim t^{1/2}$, consistent with expectations from nucleation and  growth kinetics\cite{chaikin_principles_1995} (see \textbf{Figure S\ref{SI-DLS}} and associated text). These kinetics are also consistent with Wedekind et al.'s simulations of crystal nucleation close to a colloid-liquid phase transition\cite{wedekind_optimization_2015}. They found that crystal (SL) formation in the region between the colloid-metastable liquid binodal and its spinodal curve appears to follow one-step kinetics despite the necessary presence of the liquid state; they regard this regime as exhibiting “effective one-step kinetics.” 
Alternatively, one might consider the  possibility that one- and two-step pathways co-occur with similar rates, i.e., compete with one another, in this region of the phase diagram. The “effective one-step” mechanism is, however, more likely, not only because of correspondence with this simulation work. First, we observe that the two-step pathway dominates kinetically deeper beneath the binodal (\textbf{Figure \ref{Fkinetics}d}). Second, it seems less plausible for independent one- and two-step pathways to have essentially identical rates over the entire range studied with the Avrami analysis.
Ultimately, \textbf{Figure \ref{Fkinetics}e} layers kinetic information on top of thermodynamic information to succinctly summarize the variety of behaviors accessible with this system, and can be used to prescribe conditions for diverse self-assembly outcomes.
 
Through this establishment of possible kinetic pathways, we find that the liquid phase provides new opportunities to  optimize self-assembly. First, liquid appears before or in tandem with  SL for all quenches beneath the colloid-liquid binodal in \textbf{Figure \ref{Fkinetics}e}, suggesting that the two-step and effective one-step processes are faster than the one-step process in this region, consistent with findings from simulations\cite{ten_wolde_enhancement_1997,wedekind_optimization_2015,haxton_crystallization_2015} and protein crystallization\cite{haas_interface_2000,galkin_control_2000} experiments 
(\textbf{Figure \ref{Fdesign}b}). It is still possible for the rate of the one-step process outside of the colloid-metastable coexistence liquid region to be faster than the rate of the effective one-step or two-step processes inside the colloid-metastable liquid coexistence region if larger NC volume fractions are used in the one-step process. Increasing the NC volume fraction to increase the rate of a one-step process, however, often introduces disorder (purple curve, \textbf{Figure \ref{Fdesign}c})\cite{haxton_crystallization_2015,tanner_situ_2024}. 

Second, we find that self-assembly in the liquid environment has important implications for the relationship between rate and resulting SL order. \textbf{Figure \ref{Fdesign}a} compares the $\sim$equilibrium (111) peak widths (\textbf{Figures S\ref{SI-FWHM}} and  \textbf{S\ref{SI-FWHMfit}}) as a function of volume fraction, $\phi_0$ (grey), to the corresponding trend in  rate constant, $k$, for the same set of effective one-step experiments (orange in \textbf{Figure \ref{Fkinetics}e}). We find that as the rate increases, the peak width decreases, i.e., SL quality improves (orange curve, \textbf{Figure \ref{Fdesign}c}). We have therefore identified a regime (orange region in \textbf{Figure \ref{Fkinetics}}) in which not only does the crystallization process speed up without compromising order, but the crystal quality can actually be further improved as the rate of the process increases. While we cannot unequivocally determine the mechanism of the increased SL quality, we expect it to be a combination of defect annealing (\textbf{Figure S\ref{SI-FWHM}}) and that the liquid surrounding the growing crystal surface likely reduces incorporation of defects\cite{haas_interface_2000}. Detailed kinetic study of the effective one-step regime permitted elucidation of these benefits and suggests optimal ability to maximize rate, crystallinity, and yield at and to the right of the text label in the orange region of \textbf{Figure \ref{Fkinetics}e}. The ability to crystallize from a dense liquid intermediate as compared to via diffusion-limited recruitment of particles, should be beneficial anywhere within the metastable liquid-colloid binodal. Direct comparison of the impact of one- and two-step mechanisms on SL quality at the same quench depth and volume fraction is challenging since, kinetically, one of the mechanisms tends to dominate at any specific location in the phase diagram. The liquid-containing pathways are, however, unique in that the metastable liquid provides routes to optimize the rate and quality of assembled SLs that are not present in conventional one-step crystallization and provides a powerful way to avoid gelation/kinetic trapping. This finding should be generalizable to other short-range interacting systems.

\textbf{\textit{Key factors for increasing crystallinity concomittantly with crystallization driving force.}} Combining the above insights about the interaction potential range and depth with the effective one-step kinetic pathway uncovered to achieve high rates of formation concurrently with high SL quality provides a prescription for the ultimate control of SL self-assembly, which should be translatable to other nanoscale systems. For short-range-interacting systems, the kinetic parameter space associated with the effective one-step region is determined by the location of the colloid-metastable liquid binodal, which can be controlled by the range of the interparticle interactions. Achieving short-range interactions via control of electrostatics is a highly desirable design strategy: tuning electrostatic repulsion relative to empirically determined van der Waals attraction provides access to a broad, rich parameter space in which to tune kinetic rates. It furthermore allows the yield, size, and crystallinity of self-assembly products to be controlled. Additionally, as illustrated by comparing the Au and PbS NC SL assembly, even within the context of electrostatic stabilization and short-range interacting particles, accessing the effective one-step regime to simultaneously optimize crystallization rate and product quality requires the ability to carefully tune the short-range potential.
We propose that the NC-to-solvent dielectric constant ratio, which determines the effective NC size and associated interaction potential range, is critical to tune. 
Specifically, it must be sufficiently high to achieve short-range interactions that suppress the colloid-liquid binodal beneath the colloid-SL binodal while the solvent dielectric constant is kept high enough that, upon quenching via an increase of the ionic strength, the potential depth becomes no greater than $\sim$3 $k_{\mathrm{B}}T$ and the range does not become too small, so as to avoid gelation\cite{noro_role_1999}.
This prescription is achieved with the PbS NC system using a dielectric ratio of $\sim$2, as compared to the Au system whose ratio is unbounded. A range of ratios are accessible by varying the relative proportions of miscible solvents, such as the NMF and DMF used here, so that the accessible quench potential depths that are fine-tuned by the ionic strength surround the identified few-$k_{\mathrm{B}}T$ window. Examples are shown in \textbf{Figure S\ref{SI-dielectric}}. These design principles can be applied directly to other
colloidal and protein\cite{galkin_control_2000,zhang_charge-controlled_2012} systems to test their generality and optimize their utility in promoting intricate material design on the nanoscale and also at the microscale, provided index matching is no longer necessary,\cite{noro_role_1999} for example by having a scattering technique take the place of optical microscopy. It furthermore suggests the possibility of using inorganic nanoscale systems as models for liquid-liquid phase separation of biomolecules\cite{michaels_amyloid_2023} when X-ray probes can provide additional, higher-resolution, insight that would be incompatible with proteins themselves. Furthermore, using similar \textit{in situ} measurement and analysis, predictive optimal protocols for assembly and function of a wide range of other complex systems ranging from materials for catalysis\cite{uchida_modular_2018} to drug delivery\cite{shin_covid-19_2020} to energy storage\cite{guo_self-assembly_2016} will also come into reach.

\section*{Methods}

\textbf{\textit{In situ} SAXS experiments.} SAXS data were collected at the Stanford Synchrotron Radiation Lightsource (SSRL) at beamline 1-5 with a photon energy of 15 keV and beam size of 600 $\times$ 600 $\mu$m and at the Advanced Light Source at beamline 7.3.3 with a photon energy of 10 keV and beam size of $\sim$500 $\mu$m (see \textbf{Figure S\ref{SI-aparatus}}). Stock solutions of 5.8$\pm$0.3 nm PbS NCs with $\mathrm{K}_4\mathrm{Sn}_2\mathrm{S}_6$ ligands (see Supporting Information) in N-methylformamide (NMF) and 0.5 M $\mathrm{K}_3\mathrm{AsS}_4$ salt in NMF were prepared following a procedure previously outlined\cite{coropceanu_self-assembly_2022}. In a nitrogen filled glovebox, 400-500 $\mu$Ls of a solution of 5.8 nm PbS NCs with $\mathrm{K}_4\mathrm{Sn}_2\mathrm{S}_6$ ligands in mixtures of NMF and N,N-dimethylformamide (DMF) was loaded into a 2 mm path length quartz cuvette with custom 200 $\mu$m thick windows. A small stir bar was placed into the cuvette in the plane of the cuvette and the cuvette was then sealed using a rubber septum and parafilm. A syringe preloaded with a solution of 0.5 M $\mathrm{K}_3\mathrm{AsS}_4$ in NMF was attached to the cuvette via Teflon tubing through the septum. The tubing-septum interface was sealed with epoxy. The gas-tight apparatus was carefully moved into the beam path and the syringe placed onto a New Era syringe pump (model NE-1000). X-ray scattering data were collected continuously while stirring the solution using a magnetic stirrer from Ultrafast Systems. All X-ray scattering patterns were collected using $\sim$0.1$-$1 s exposures at a rate of one pattern every $\sim$0.1$-$5 s. For each \textit{in situ} experiment, after about 5 min of data acquisition, the excess salt in NMF solution was injected using the syringe pump at a rate of $\sim$1 mL/s. The injection took $\sim$ 1 - 10 s depending on how much salt was added. The ionic strength of the solution post-quench is $I=\frac{1}{2}\sum_{n=1}^{N}c_nz_n^2$, where $c_n$ is the concentration of ion species $n$ in molar, $z_n$ is the valency of ion $n$, and $N$ is the number of different ion species in solution. The Debye length of the solution post-quench is $\lambda=\sqrt{\frac{\epsilon_r \epsilon_0 k_\mathrm{B}T}{\sum_{n=1}^{N}c_nz_n^2e^2}}$ where $\epsilon_r$ is the solvent dielectric constant of 88, $\epsilon_0$ is the vacuum permittivity, and $e$ is the elemental charge. The apparatus was kept at room temperature (see Supporting Information for additional temperature considerations). Data were continuously acquired after injection for up to two hours. SAXS patterns of cuvettes filled with NMF/DMF and varying amounts of $\mathrm{K}_3\mathrm{AsS}_4$ salt were taken for background subtraction (\textbf{Figure S\ref{SI-background}}). 

\textbf{Modeling of SAXS patterns.} We model the background-subtracted scattered intensity as $I(q) = I_\mathrm{colloid}(q) + I_\mathrm{liquid}(q) + I_\mathrm{SL}(q) + I_\mathrm{LQ}(q)$, where $I_\mathrm{colloid}(q)$ is the scattered intensity from dilute, polydisperse, hard spheres, $I_\mathrm{liquid}(q)$ is the scattered intensity at high $q$ from a liquid of hard spheres, $I_\mathrm{SL}(q)$ is the scattered intensity at high $q$ from \textit{fcc} SLs, and $I_\mathrm{LQ}(q)$ is the scattering at low $q$ due to the finite size of liquid and SL clusters. For $I_\mathrm{colloid}(q)$, $I_\mathrm{SL}(q)$, and $I_\mathrm{LQ}(q)$, we use models previously developed\cite{tanner_situ_2024}. To compute $I_\mathrm{liquid}(q)$, we solve the Ornstein-Zernicke equation using the Percus-Yevick closure relation for the case of hard spheres\cite{percus_analysis_1958,wertheim_exact_1963}. To explicitly compute $I_\mathrm{colloid}(q)$ and $I_\mathrm{liquid}(q)$ we use xrsdkit (https://github.com/scattering-central/xrsdkit). See supporting information for further details on the components of the model. 

\textbf{Calculation of fraction of NCs in each phase.} The fraction of NCs in phase $P$ at a given time, $f_P(t)$, is approximated by $f_P(t)=Q_{\mathrm{lim},P}(t)/(Q_{\mathrm{lim,colloid}}(t)+Q_{\mathrm{lim,liquid}}(t)+Q_{\mathrm{lim,SL}}(t))$. Here, $Q_{\mathrm{lim},P}(t)$\cite{glatter_small_1982} is the limited range scattering invariant,  $Q_{\mathrm{lim},P}(t)=\int_{q_{min}}^{q_{max}}q^2I_P(q,t)dq$, where $I_P(q,t)$ is the contribution of phase $P$ in the fit of the model to the SAXS pattern at time $t$, $P$ is colloid, liquid, or SL, and $q_\mathrm{min}$ and $q_\mathrm{max}$ are the minimum and maximum values for the $q$ range of the data we fit ($\sim$0.005-0.25 $\text{\AA}^{-1}$). Owing to the limited $q$ range, this calculation underestimates the invariant, especially for the liquid and SL phases, but it does not significantly impact the kinetics observed (\textbf{Figure S\ref{SI-invariant}}).

\textbf{Avrami analysis.} We analyzed kinetics of liquid and SL formation using the Avrami equation, \cite{cantor_avrami_2020}  $V(t)=1-\exp(kt^n)$, where $V(t)$ is the volume fraction of the new phase. The Avrami exponent, $n$, which is typically between 1 and 4, describes the dimensionality of the system, with a contribution of 1 for nucleation and a contribution of 1 for each of the growth dimensions. We assume homogeneous nucleation and growth in 3 dimensions and thus expect an Avrami exponent of $n\sim 4$. For $n=4$, the Avrami rate constant, $k$, is $k=\frac{1}{3}k_nk_g^3$, where $k_n$ is the nucleation rate and $k_g$ is the growth rate in a single dimension. 
We use the fractions of NCs in the liquid and SL phases obtained from the scattering invariant calculation described above as proxies for  $V(t)$, and fit these to the Avrami equation to extract $k$. Including density dependencies in the nucleating barrier and particle collision frequencies, classical nucleation theory predicts that $k$ depends on $\phi_0$ as $k=c\phi_0^3\exp(-\frac{a}{(\ln(\phi_0/\phi_\mathrm{colloid}))^2})$. \cite{debenedetti_metastable_1996,nanev_7_2015,uwaha_8_2015} The parameters $a$, which depends on temperature and surface energy of the interface of the nucleus and the colloidal phase, $c$, which arises from the growth rate and the kinetic prefactor to the nucleation rate, and $\phi_\mathrm{colloid}$, which is the volume fraction of the colloidal phase at equilibrium, do not vary with $\phi_0$. See supporting information for further details.  

\textbf{Effective size of NCs.} We estimate the effective size of the NCs by their hydrodynamic diameter and their center to center distance in the SL phase. The hydrodynamic diameter of 6.4 nm was measured via dynamic light scattering (DLS). We calculate the center-to-center distance between NCs in the SL phase directly from the location of the SL $(111)$ peak, $q_{111}$, using center-to-center distance = $=2\pi\sqrt{3}/\sqrt{2}q_{111}$. The center-to-center distances based on the range of $q_{111}$ values are $\sim 6.7-7.8$ nm.

\section*{Conflicts of Interest}
There are no conflicts to declare.

\section*{Acknowledgments}

We thank M. Delor for help designing the DLS setup. This work was supported by the Office of Basic Energy Sciences (BES), US Department of Energy (DOE) (award no. DE-SC0019375). Use of the Stanford Synchrotron Radiation Lightsource, SLAC National Accelerator Laboratory, is supported by the DOE, Office of Science, Office of Basic Energy Sciences (contract no. DE-AC02-76SF00515). Use of beamline 7.3.3 at the Advanced Light Source, Lawrence Berkeley National Laboratory is supported by the DOE, Office of Science, Office of Basic Energy Sciences (contract no. DE-AC02-05CH11231). C.P.N.T., V.R.K.W., and R.B.W. were supported by the NSF Graduate Research Fellowship. L.M.H. and J.A.T. acknowledge a National Defense Science and Engineering Graduate Fellowship. J.K.U. was supported by an Arnold O. Beckman Postdoctoral Fellowship in Chemical Sciences from the Arnold and Mabel Beckman Foundation. A.D. and L.M.H. were supported by Philomathia Graduate Student Fellowships from the Kavli Energy NanoScience Institute at UC Berkeley. D.T.L. was supported by an Alfred P. Sloan Research Fellowship. N.S.G. was supported by a David and Lucile Packard Foundation Fellowship for Science and Engineering and Camille and a Henry Dreyfus Teacher-Scholar Award. 

\begin{suppinfo}

Experimental details, modeling and simulation of SAXS patterns, additional quantitative phase diagram, scattering invariant and Avrami analysis details, SL (111) FWHM, \textit{in situ} DLS measurements, comparison of Au and PbS NC phase behavior. 

\end{suppinfo}
\bibliography{PbS_paper_2023.bib}


\end{document}